\documentclass[useAMS]{mn2e}

\usepackage{latexsym,graphicx}

\usepackage{color}

\newcommand\kms{{\rm\,km\,s^{-1}}}
\newcommand\msun{\rm\,M_\odot}

\newcommand{\MC}{\multicolumn}

\def\apgt{\ {\raise-.5ex\hbox{$\buildrel>\over\sim$}}\ }
\def\aplt{\ {\raise-.5ex\hbox{$\buildrel<\over\sim$}}\ }

\title[Discovery of two new Galactic candidate luminous blue variables with WISE]{Discovery of two new Galactic
candidate luminous blue variables with WISE\footnotemark[0]\thanks{%
Based on observations obtained with the South African Large
Telescope (SALT), commissioning programs 2010-1-RSA\_OTH-001 and
2010-1-RSA\_OTH-013.} }
\author[V.V.Gvaramadze et al.]
       {V. V.~Gvaramadze,$^{1,2}$\thanks{E-mail: vgvaram@mx.iki.rssi.ru}
        A. Y.~Kniazev,$^{3,4,1}$ A. S.~Miroshnichenko,$^{5}$
       \newauthor
       L. N.~Berdnikov,$^{1,2}$ N.~Langer,$^{6}$ G. S.~Stringfellow,$^{7}$ H.~Todt,$^{8}$ W.-R.~Hamann,$^{8}$
       \newauthor
       E. K.~Grebel,$^{9}$ D.~Buckley,$^{4}$ L.~Crause,$^{3}$ S.~Crawford,$^{3,4}$ A.~Gulbis,$^{3,4}$
       \newauthor
       C.~Hettlage,$^{3,4}$ E.~Hooper,$^{10}$ T.-O.~Husser,$^{11}$ P.~Kotze,$^{3,4}$ N.~Loaring,$^{3,4}$
       \newauthor
       K. H.~Nordsieck,$^{10}$ D.~O'Donoghue,$^{4}$ T.~Pickering,$^{3,4}$
       S.~Potter,$^{3}$ E.~Romero
       \newauthor
       Colmenero,$^{3,4}$ P.~Vaisanen,$^{3,4}$ T.~Williams,$^{12}$ M.~Wolf,$^{10}$ D. E.~Reichart,$^{13}$
       \newauthor
        K. M.~Ivarsen,$^{13}$ J. B.~Haislip,$^{13}$ M.C.~Nysewander$^{13}$ and A. P.~LaCluyze$^{13}$ \\
        $^{1}$Lomonosov Moscow State University, Sternberg Astronomical Institute, Universitetskij Pr. 13, Moscow 119992, Russia\\
        $^{2}$Isaac Newton Institute of Chile, Moscow Branch, Universitetskij Pr. 13, Moscow 119992, Russia\\
        $^{3}$South African Astronomical Observatory, PO Box 9, 7935 Observatory, Cape Town,
        South Africa \\
        $^{4}$Southern African Large Telescope Foundation, PO Box 9, 7935 Observatory, Cape Town,
        South Africa \\
        $^{5}$ Department of Physics and Astronomy, University of North Carolina at Greensboro, Greensboro, NC 27402-6170, USA \\
        $^{6}$ Argelander-Institut f\"ur Astronomie der Universit\"at Bonn, Auf dem H\"ugel 71, 53121, Bonn, Germany \\
        $^{7}$ CASA, University of Colorado, UCB 389, Boulder, CO 80309, USA \\
        $^{8}$ Institute for Physics and Astronomy, University of Potsdam, 14476 Potsdam, Germany \\
        $^{9}$ Astronomisches Rechen-Institut, Zentrum f\"{u}r Astronomie der Universit\"{a}t Heidelberg, M\"{o}nchhofstr. 12-14, 69120 Heidelberg, \\
        Germany \\
        $^{10}$Department of Astronomy, University of Wisconsin-Madison, 475 N. Charter St., Madison, WI 53706, USA \\
        $^{11}$Institut f\"{u}r Astrophysik Georg-August-Universit\"{a}t, Friedrich-Hund-Platz 1, 37077 Gottingen, Germany \\
        $^{12}$Department of Physics and Astronomy, Rutgers University, 136 Frelinghuysen Road, Piscataway, NJ 08854, USA \\
        $^{13}$Department of Physics and Astronomy, University of North Carolina, Chapel Hill, NC 27599, USA
        }
\begin{document}

\date{Accepted 2012 January 13. Received 2012 January 13; in original form 2011 December 5}


\maketitle

\label{firstpage}

\begin{abstract}
We report the discovery of two new Galactic candidate luminous
blue variable (cLBV) stars via detection of circular shells
(typical of known confirmed and cLBVs) and follow-up spectroscopy
of their central stars. The shells were detected at $22 \, \mu$m
in the archival data of the Mid-Infrared All Sky Survey carried
out with the Wide-field Infrared Survey Explorer (WISE). Follow-up
optical spectroscopy of the central stars of the shells conducted
with the renewed Southern African Large Telescope (SALT) showed
that their spectra are very similar to those of the well-known
LBVs P\,Cygni and AG\,Car, and the recently discovered cLBV MN112,
which implies the LBV classification for these stars as well. The
LBV classification of both stars is supported by detection of
their significant photometric variability: one of them brightened
in the $R$- and $I$-bands by $0.68\pm0.10$ mag and $0.61\pm0.04$
mag, respectively, during the last 13-18 years, while the second
one (known as Hen 3-1383) varies its $B,V,R,I$ and $K_{\rm s}$
brightnesses by $\simeq 0.5-0.9$ mag on time-scales from 10 days
to decades. We also found significant changes in the spectrum of
Hen 3-1383 on a timescale of $\simeq 3$ months, which provides
additional support for the LBV classification of this star.
Further spectrophotometric monitoring of both stars is required to
firmly prove their LBV status. We discuss a connection between the
location of massive stars in the field and their fast rotation,
and suggest that the LBV activity of the newly discovered cLBVs
might be directly related to their possible runaway status.
\end{abstract}


\section{Introduction}
\label{sec:intro}

Massive stars go through a sequence of transitional evolutionary
stages whose duration and order are still not well established
(Chiosi \& Maeder 1986; Langer et al. 1994; Stothers \& Chin 1996;
Maeder \& Meynet 2011). Of these stages, perhaps the most
important in the evolutionary sense and interesting in
observational manifestations is the Luminous Blue Variable (LBVs)
stage (Conti 1984; Wolf 1992; Humphreys \& Davidson 1994; van
Genderen 2001; Vink 2009). During this stage a massive star could
exhibit irregular, strong spectroscopic and photometric
variability on timescales from years to decades or longer. This
variability manifests itself in changes of the stellar spectrum
from that of late O/early B-type supergiants to that of A/F-type
ones (e.g. Stahl \& Wolf 1982; Stahl et al. 2001) and in the
brightness increase by several magnitudes during the cooler
(outburst) state. Major eruptions of LBV stars can masquerade as
peculiar supernovae (Goodrich et al. 1989; Filippenko et al. 1995;
Van Dyk et al. 2002; Smith et al. 2011), while some LBVs could
perhaps be the direct progenitors of core-collapse supernovae
(Kotak \& Vink 2006; Smith et al. 2007; Gal-Yam \& Leonard 2009).

The spectrophotometric variability of LBVs is often accompanied by
drastic changes in the stellar mass-loss rate so that essentially
all LBVs are surrounded by compact ($\sim 0.1-1$ pc in diameter)
shells (visible in the infrared, optical and/or radio wavebands)
with a wide range of morphologies, ranging from circular to
bipolar and triple-ring forms (e.g. Nota et al. 1995; Weis 2001;
Smith 2007; Gvaramadze, Kniazev \& Fabrika 2010c). The origin of
the LBV shells can be attributed either to the interaction of
stellar winds during the subsequent phases of evolution of massive
stars (e.g. Robberto et al. 1993; Garcia-Segura, Langer, Mac Low
1996) or to instantaneous ejections (e.g. Smith \& Owocki 2006)
caused by some catastrophic events (triggered either by internal
or external factors).

The LBV phenomenon is still poorly understood, which is mostly due
to the scarcity of known LBV stars (caused by the short duration,
a few $10^3 -10^4$ years, of the LBV phase). Several years ago
there were only 12 confirmed Galactic members of the class and 24
candidate LBVs (cLBVs) (Clark, Larionov \& Arkharov 2005; Vink
2009). The discovery of additional Galactic LBVs would, therefore,
greatly advance the field and help in understanding the
evolutionary status and connection of LBVs to other massive
transient stars, the diversity of the environments in which they
form, and their distribution in the Galaxy. One can also expect
that the increase of the number of known LBVs would allow us to
understand the driving mechanism(s) of the LBV phenomenon and to
answer the question: whether a deficiency of LBVs with
luminosities between $\log (L/L_\odot )\simeq 5.6$ and 5.8 is real
or is a result of the small-number statistics (Smith, Vink \& de
Koter 2004)?

A detection of LBV-like shells can be considered as an indication
that their associated stars are massive and evolved (e.g. Bohannan
1997; Clark et al. 2003), and therefore could be used for
selection of candidate massive stars for follow-up spectroscopy.
Because of the huge interstellar extinction in the Galactic plane,
the most effective channel for the detection of (circumstellar)
shells is through imaging with modern infrared telescopes. The
infrared imaging could also be crucial for revealing those LBV
stars which are presently enshrouded in dusty envelopes created by
major eruptions in the recent past and therefore are heavily
obscured in the optical band. Application of this approach using
the archival data of the {\it Spitzer Space Telescope} resulted in
detection of dozens of new cLBV, blue supergiant, and Wolf-Rayet
(WR) stars in the Milky Way (Gvaramadze et al. 2009, 2010a,b;
Gvaramadze, Kniazev \& Fabrika 2010c; Wachter et al. 2010, 2011;
Mauerhan et al. 2010; Stringfellow et al. 2012a,b). The majority
of these shells are visible only in 24\,$\mu$m images obtained
with the Multiband Imaging Photometer for {\it Spitzer} (MIPS;
Rieke et al. 2004) and most of them were detected in the archival
data of the 24 and 70 Micron Survey of the Inner Galactic Disc
with MIPS (MIPSGAL; Carey et al. 2009), which mapped 278 square
degrees of the inner Galactic plane ($|b|<1\degr$ is covered for
$5\degr < l < 63\degr$ and $298\degr < l <355\degr$ and
$|b|<3\degr$ is covered for $|l|<5\degr$). The high angular
resolution of the MIPS images (6 arcsec at 24\,$\mu$m) even
allowed us to detect a new circular shell in the Large Magellanic
Cloud (LMC) and therethrough to discover a new WR star -- the
first-ever extragalactic evolved massive star detected with {\it
Spitzer} (Gvaramadze et al. 2011a; Gvaramadze et al., in
preparation).

Due to a limited sky coverage of the MIPSGAL and other {\it
Spitzer} Legacy
Programs\footnote{http://irsa.ipac.caltech.edu/Missions/spitzer.html},
however, many evolved massive stars with circumstellar shells
might remain undetected. The situation was somewhat improved with
the recent data release of the Mid-Infrared All Sky Survey carried
out with the Wide-field Infrared Survey Explorer (WISE; Wright et
al. 2010)\footnote{The current release of the WISE data covers 57
per cent of the sky.}. This survey provides images at four
wavelengths: 3.4, 4.6, 12 and 22\,$\mu$m, with an angular
resolution of 6.1, 6.4, 6.5 and 12.0 arcsec, respectively, and
allows a search for infrared circumstellar shells in the regions
not covered by {\it Spitzer}. The factor of two lower angular
resolution of the 22\,$\mu$m WISE data, however, makes many of the
shells discovered with {\it Spitzer} almost indiscernible, so that
we expect that only the most extended yet unknown shells would be
detected with WISE.

In this paper, we report the discovery of two new Galactic cLBVs
via detection of circular shells with WISE and follow-up optical
spectroscopy of their central stars with the renewed Southern
African Large Telescope (SALT). The new shells and their central
stars are presented in Section\,\ref{sec:shells}.
Section\,\ref{sec:obs} describes our spectroscopic follow-up of
the central stars and the data reduction. Archival and
contemporary photometry of the stars is presented in
Section\,\ref{sec:phot}. The obtained spectra are discussed in
Section\,\ref{sec:spec}. The discussion is given in
Section\,\ref{sec:dis}. We summarize and conclude in
Sect.\,\ref{sec:sum}.

\section{New circular shells and their central stars}
\label{sec:shells}

Two circular shells, which are the main subject of this paper,
were discovered using the WISE 22\,$\mu$m archival data. Both
shells contain central stars (see Table\,\ref{tab:det} for their
coordinates), which are prominent not only in all four WISE
wavebands but also in the optical range. We call these shells WS1
and WS2, i.e. the WISE shells, and, in that follows, we will use
for the central stars the same names as for the associated
nebulae.

\begin{figure*}
\begin{center}
\includegraphics[width=18cm,angle=0,clip=]{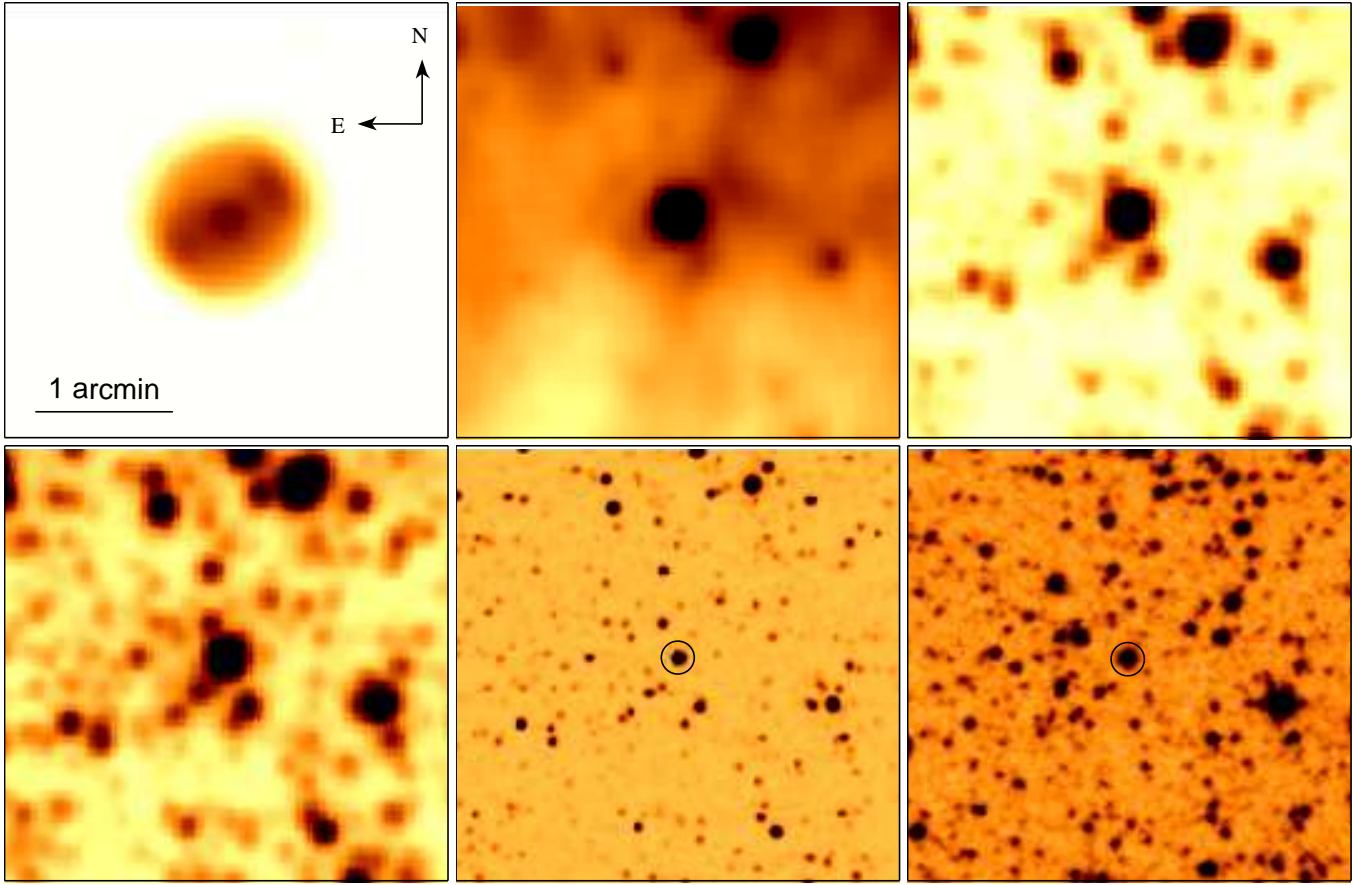}
\end{center}
\caption{From left to right, and from top to bottom: WISE 22, 12,
4.6 and 3.4\,$\mu$m, 2MASS $K_{\rm s}$-band and DSS-II red band
images of the circular shell WS1 and its central star (marked by a
circle on the 2MASS and DSS-II images). The orientation and the
scale of the images are the same.
    }
\label{fig:shell-1}
\end{figure*}

Figure\,\ref{fig:shell-1} shows the WISE, 2MASS $K_{\rm s}$-band
(Skrutskie et al. 2006) and DSS-II red band (McLean et al. 2000)
images of the field containing WS1. At 22\,$\mu$m the shell is
almost circular with a radius of $\simeq 35$ arcsec. One can also
see an elongated (toroid-like?) region of enhanced brightness
stretched within the shell in the northwest-southeast direction.
The 12\,$\mu$m image shows the gleam of emission probably
associated with WS1 and an arc-like structure faced towards the
centre of the 22\,$\mu$m shell. The brightest part of this arc
coincides with the region of enhanced brightness at the northwest
of the shell. The central star of WS1 corresponds to the 2MASS
J13362862-6345387 source (Skrutskie et al. 2006).

\begin{figure*}
\begin{center}
\includegraphics[width=18cm,angle=0,clip=]{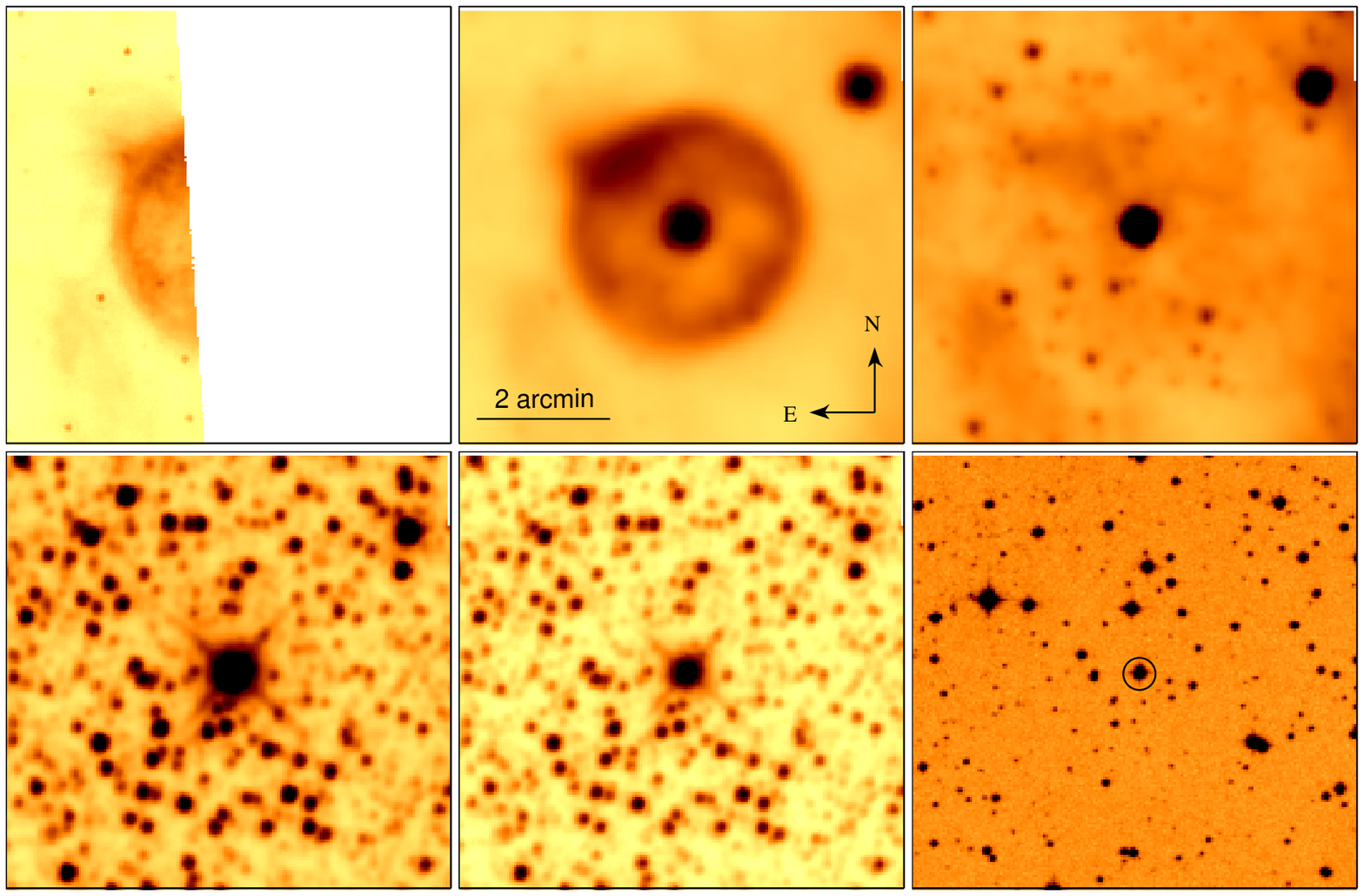}
\end{center}
\caption{From left to right, and from top to bottom: {\it Spitzer}
MIPS $24\,\mu$m, WISE 22, 12, 4.6 and $3.4\,\mu$m and DSS-II red
band images of the circular shell WS2 and its central star (marked
by a circle on the DSS-II image). The MIPS image is truncated
because the MIPSGAL survey covers only a part of the shell. The
orientation and the scale of the images are the same.
    }
\label{fig:shell-2}
\end{figure*}

WS2 was discovered during our search for bow shocks generated by
OB stars running away from the massive star-forming region
NGC\,6357 (for motivation and the results of this search see
Gvaramadze \& Bomans 2008 and Gvaramadze et al. 2011b,
respectively). A by-product of the search, which used the archival
data of the {\it Midcourse Space Experiment} ({\it MSX}) satellite
(Price et al. 2001) and the MIPSGAL and WISE surveys, is the
discovery of three LBV-like shells, one of which, WS2, has a
central star visible in the optical region. Unfortunately, the
MIPSGAL survey covers only a part of this quite extended (with a
radius of $\simeq 115$ arcsec) shell, so that the full extent of
WS2 was revealed only at 22\,$\mu$m with WISE (see
Fig.\,\ref{fig:shell-2}). The MIPS 24\,$\mu$m and the WISE
22\,$\mu$m images of WS2 have already been presented in Gvaramadze
et al. (2011b).

WS2 has a clear circular limb-brightened structure with a region
of reduced brightness in the southeast of the shell and a
chimney-like blow-up (more prominent at 24\,$\mu$m due to the
better angular resolution of the MIPS image) attached to a region
of enhanced brightness in the northeast. The WISE 12\,$\mu$m image
shows the gleam of emission possibly related to WS2. Particularly,
one can see two wings apparently extending from the central star
towards the shell. One of them connects the star to the brightest
part of the shell and then diverges to the ``chimney", while the
second (less prominent) wing connects the star with the west rim
of the shell. Although it is not clear whether or not these wings
are related to WS2, we note that similar structures can also be
seen within the shell of the cLBV MN96 (see Fig.\,1(n) in
Gvaramadze et al. 2010c or Fig.\,1 in Stringfellow et al. 2012a).

Using the SIMBAD
database\footnote{http://simbad.u-strasbg.fr/simbad/} and the
VizieR catalogue access
tool\footnote{http://webviz.u-strasbg.fr/viz-bin/VizieR}, we
identified WS2 with the emission-line star Hen\,3-1383, classified
in the literature as an M3e (The 1961), an Mep? (Sanduleak \&
Stephenson 1973; Allen 1978) and a candidate symbiotic (Henize
1976) star. Hen 3-1383 is also known as a variable source of
thermal radio emission (Seaquist \& Taylor 1990; Altenhoff, Thum
\& Wendker 1994).

The morphology of the WS1 and WS2 shells is very similar to that
of the majority of circumstellar shells associated with known
(c)LBVs and related evolved massive stars (see Gvaramadze et al.
2010c), which allowed us to expect that the central stars of the
shells are evolved massive stars as well. Follow-up spectroscopy
of these stars confirmed our expectation and leads to the
discovery of two new cLBVs.

\section{Observations}

\subsection{Spectroscopic follow-up}
\label{sec:obs}

\begin{table}
\centering{ \caption{Equivalent widths (EWs), FWHMs and radial
heliocentric velocities (RVs) of some detected lines in the
spectrum of WS1 (taken on 2011 June 12). RVs of lines noticeably
affected by P\,Cygni absorptions are starred.} \label{tab:ws1}
\begin{tabular}{lccc} \hline
\rule{0pt}{10pt} $\lambda_{0}$(\AA) Ion  & EW($\lambda$)  &
FWHM($\lambda$)  &    RV          \\
 & (\AA) & (\AA) & ($\kms$) \\
\hline
4340\ H$\gamma$\        &  4.7$\pm$0.2  &  4.97$\pm$0.21  &    83$\pm$6*    \\
4471\ He\ {\sc i}\      &  3.4$\pm$0.3  &  8.08$\pm$0.79  &  205$\pm$22*  \\
4554\ Si\ {\sc iii}\    &  0.5$\pm$0.1  &  5.18$\pm$0.72  &  198$\pm$20*  \\
4631\ N\ {\sc ii}\      &  0.2$\pm$0.1  &  6.16$\pm$0.22  &  102$\pm$5    \\
4658\ [Fe\ {\sc iii}]\  &  0.12$\pm$0.04 & 6.29$\pm$0.42 &  -103$\pm$8    \\
4686\ He\ {\sc ii}\    &  0.4$\pm$0.1  & 10.82$\pm$0.43  &    94$\pm$8    \\
4713\ He\ {\sc i}\      &  1.1$\pm$0.1  &  5.82$\pm$0.19  &    90$\pm$5    \\
4861\ H$\beta$\        & 16.3$\pm$0.3  &  6.53$\pm$0.12  &    42$\pm$3*    \\
4922\ He\ {\sc i}\      &  1.3$\pm$0.2  &  4.52$\pm$0.63  &  124$\pm$16*  \\
5016\ He\ {\sc i}\      &  4.3$\pm$0.2  &  5.98$\pm$0.27  &    79$\pm$7*    \\
5128\ Fe\ {\sc iii}\    &  0.4$\pm$0.1  &  4.25$\pm$1.20  &  139$\pm$30*  \\
5156\ Fe\ {\sc iii}\    &  0.2$\pm$0.1  &  3.33$\pm$1.59  &  136$\pm$39*  \\
5755\ [N\ {\sc ii}]\    &  0.8$\pm$0.2  &  8.14$\pm$0.72  &  -41$\pm$16    \\
5834\ Fe\ {\sc iii}\    &  0.5$\pm$0.2  &  6.65$\pm$0.37  &    -1$\pm$8    \\
5876\ He\ {\sc i}\      & 16.5$\pm$0.5  &  6.29$\pm$0.19  &    64$\pm$4*    \\
5979\ Fe\ {\sc iii}\    &  0.7$\pm$0.2  &  6.63$\pm$0.21  &  -16$\pm$4    \\
6000\ Fe\ {\sc iii}\    &  0.2$\pm$0.1  &  4.06$\pm$0.54  &    -6$\pm$12    \\
6033\ Fe\ {\sc iii}\    &  0.7$\pm$0.2  &  6.27$\pm$0.10  &  -14$\pm$2    \\
6347\ Si\ {\sc ii}\    &  1.3$\pm$0.2  & 12.37$\pm$0.49  &  -56$\pm$9    \\
6371\ Si\ {\sc ii}\    &  0.5$\pm$0.1  &  6.84$\pm$0.23  &  -31$\pm$5    \\
6482\ N\ {\sc ii}\      &  1.0$\pm$0.1  &  5.98$\pm$0.24  &    76$\pm$5*    \\
6563\ H$\alpha$\        & 63.2$\pm$0.9  &  7.86$\pm$0.12  &    32$\pm$2*    \\
6611\ N\ {\sc ii}\      &  0.7$\pm$0.1  &  9.13$\pm$0.49  &  -42$\pm$8    \\
6678\ He\ {\sc i}\      &  6.9$\pm$0.3  &  6.27$\pm$0.28  &    69$\pm$5*    \\
7065\ He\ {\sc i}\      & 10.7$\pm$0.3  &  7.20$\pm$0.17  &    30$\pm$3*    \\
\hline
\end{tabular}
}
\end{table}
\begin{table*}
\centering{ \caption{Equivalent widths (EWs), FWHMs and radial
heliocentric velocities (RVs) of some detected lines in the
spectra of WS2 taken on 2011 June 13 (1) and 2011 September 10
(2). RVs of lines noticeably affected by P\,Cygni absorptions are
starred.} \label{tab:ws2}
\begin{tabular}{lrrrrrrrrr} \hline
\rule{0pt}{10pt}
$\lambda _{0}$(\AA) Ion && \MC{2}{c}{EW($\lambda$)}          &&  \MC{2}{c}{FWHM($\lambda$)}      &&  \MC{2}{c}{RV}      \\
              && \MC{2}{c}{(\AA)}                  &&  \MC{2}{c}{(\AA)}                &&  \MC{2}{c}{($\kms$)} \\ \hline
              && \MC{1}{c}{(1)}&\MC{1}{c}{(2)}    &&  \MC{1}{c}{(1)}&\MC{1}{c}{(2)}  &&  \MC{1}{c}{(1)}&\MC{1}{c}{(2)} \\
\hline
4340\ H$\gamma$\      &&  6.92$\pm$0.35  &  8.88$\pm$0.49  &&  6.76$\pm$0.40  & 6.23$\pm$0.40  &&  129$\pm$11* & 107$\pm$12*  \\
4471\ He\ {\sc i}\    &&  3.63$\pm$0.21  &  2.00$\pm$0.07  &&  7.01$\pm$0.47  & 6.61$\pm$0.31  &&  134$\pm$13* & 76$\pm$8*    \\
4554\ Si\ {\sc iii}\  &&  2.66$\pm$0.10  &  1.52$\pm$0.28  &&  9.23$\pm$0.27  & 8.13$\pm$1.96  &&  54$\pm$8    & 44$\pm$50    \\
4631\ N\ {\sc ii}\    &&  1.41$\pm$0.05  &  0.66$\pm$0.15  &&  8.08$\pm$0.31  & 7.97$\pm$0.49  &&  76$\pm$7    & 63$\pm$10    \\
4713\ He\ {\sc i}\    &&  1.98$\pm$0.05  &  2.20$\pm$0.09  &&  7.22$\pm$0.13  & 7.69$\pm$0.35  &&  96$\pm$3    & 85$\pm$9    \\
4861\ H$\beta$\        && 32.69$\pm$0.56 & 31.61$\pm$0.49 &&  7.08$\pm$0.14  & 7.03$\pm$0.12 && 72$\pm$4* & 58$\pm$3*   \\
4922\ He\ {\sc i}\    &&  4.22$\pm$0.38  &  --            &&  6.21$\pm$0.64  & --            &&  164$\pm$17* & --        \\
5016\ He\ {\sc i}\    &&  7.90$\pm$0.45  &  10.71$\pm$0.58 &&  8.34$\pm$0.56  & 8.32$\pm$0.52  && 176$\pm$14*  & 179$\pm$13*  \\
5128\ Fe\ {\sc iii}\  &&  1.36$\pm$0.10  &  1.45$\pm$0.11  &&  5.72$\pm$0.42  & 6.15$\pm$0.47  && 55$\pm$11*  & 77$\pm$12*  \\
5156\ Fe\ {\sc iii}\  &&  1.72$\pm$0.07  &  2.12$\pm$0.10  &&  6.11$\pm$0.23  & 6.52$\pm$0.33  && 75$\pm$6*    & 91$\pm$8*    \\
5171\ Fe\ {\sc ii}\    &&  3.25$\pm$0.09  &  3.55$\pm$0.10  &&  7.64$\pm$0.21  & 8.29$\pm$0.26  && $-36\pm$5*  & $-26\pm$6*    \\
5319\ Fe\ {\sc ii}\    &&  2.41$\pm$0.10 &  2.61$\pm$0.09  &&  8.55$\pm$0.20 & 9.82$\pm$0.54  && $-56\pm$5    & $-65\pm$8    \\
5334\ [Fe\ {\sc ii}]\  &&  0.70$\pm$0.11 &  0.65$\pm$0.09  && 11.63$\pm$0.90 & 9.89$\pm$1.44  && $-24\pm$18  & $-40\pm$32  \\
5376\ [Fe\ {\sc ii}]\  &&  0.58$\pm$0.09 &  0.64$\pm$0.06  && 13.56$\pm$0.32 & 11.25$\pm$0.64 && $-16\pm$7    & 7$\pm$15    \\
5755\ [N\ {\sc ii}]\  &&  2.07$\pm$0.09  &  1.98$\pm$0.08  && 13.11$\pm$0.40  & 11.24$\pm$0.48 && $-122\pm$18  & $-117\pm$10  \\
5834\ Fe\ {\sc iii}\  &&  0.72$\pm$0.08  &  0.59$\pm$0.19  &&  8.72$\pm$0.23  & 4.81$\pm$1.69  && 21$\pm$5*    & $-11\pm$37*  \\
5876\ He\ {\sc i}\    && 13.04$\pm$0.26  &  16.42$\pm$0.35 &&  6.75$\pm$0.15  & 7.06$\pm$0.17  && 30$\pm$3*    & 42$\pm$4*    \\
5979\ Fe\ {\sc iii}\  &&  1.70$\pm$0.09  &  --            && 10.93$\pm$0.23  & --            &&  $-53\pm$5  & --        \\
6000\ Fe\ {\sc iii}\  &&  0.29$\pm$0.06  &  --            &&  7.51$\pm$0.65  & --            &&  $-92\pm$13  & --        \\
6033\ Fe\ {\sc iii}\  &&  0.54$\pm$0.05  &  --            &&  6.96$\pm$0.15  & --            &&  $-65\pm$3  & --        \\
6347\ Si\ {\sc ii}\    &&  1.43$\pm$0.06  &  1.29$\pm$0.05  &&  9.73$\pm$0.27  & 10.37$\pm$0.30 && 13$\pm$5*    & 42$\pm$5*    \\
6371\ Si\ {\sc ii}\    &&  0.68$\pm$0.05  &  0.66$\pm$0.04  &&  7.45$\pm$0.12  & 7.49$\pm$0.22  && $-38\pm$2    & $-1\pm$4    \\
6482\ N\ {\sc ii}\    &&  1.48$\pm$0.20 &  1.50$\pm$0.20  &&  9.26$\pm$1.34 & 9.49$\pm$1.39  && 188$\pm$26*  & 201$\pm$27*  \\
6563\ H$\alpha$\      &&136.00$\pm$2.55  & 108.98$\pm$2.10 && 11.52$\pm$0.25  & 13.73$\pm$0.30 && 43$\pm$5*    & 55$\pm$6*    \\
6678\ He\ {\sc i}\    &&  7.83$\pm$0.45  &  8.37$\pm$0.46  &&  7.77$\pm$0.51  & 7.96$\pm$0.50  && 54$\pm$10*  & 68$\pm$10*    \\
7065\ He\ {\sc i}\    && 12.28$\pm$0.44  &  --            &&  9.17$\pm$0.38  & --            &&    11$\pm$7* & --        \\
7155\ [Fe\ {\sc ii}]\  &&  1.92$\pm$0.16 &  --            && 16.82$\pm$0.86 & --            &&  $-64\pm$15  & --        \\
\hline
\end{tabular}
}
\end{table*}

Spectral observations of WS1 and WS2 were conducted with SALT
(Buckley, Swart \& Meiring 2006; O'Donoghue et al. 2006) on 2011
June 12 and 13 during the Performance Verification phase of the
Robert Stobie Spectrograph (RSS) (Burgh et al. 2003; Kobulnicky et
al. 2003). The long-slit spectroscopy mode of the RSS was used
with a 1 arcsec slit width for observations of both stars. The
grating GR900 was used to cover the spectral range of $4200-7250$
\AA \, with a final reciprocal dispersion of $0.97$ \AA \,
pixel$^{-1}$ and FWHM spectral resolution of $4.74\pm 0.05$ \AA.
The seeing during the observations was 2.1 arcsec for WS1 and 1.6
arcsec for WS2. Spectra of a Xe comparison arc were obtained to
calibrate the wavelength scale. WS1 was observed with $3\times
300$ s exposures, while WS2 with $2\times 120$ s and $2\times 600$
s ones. Because of significant $K_{\rm s}$-band photometric
variability of WS2 detected in the archival databases (see
Section\,\ref{sec:phot}), we decided to re-observe this star to
search for possible changes in its spectrum. A second spectrum of
WS2 was taken on 2011 September 10 with $3\times 800$ s exposures.
We used the same grating to cover the spectral range of
$3900-7000$ \AA \,, which allowed us to fill gaps in the
wavelength coverage of the first spectrum. The seeing during this
observation was variable, $\simeq 3$ arcsec or worse.
Spectrophotometric standard stars were observed after observation
of each object for relative flux calibration.

Primary reduction of the data was done with the SALT science
pipeline (Crawford et al. 2010). After that, the bias and gain
corrected and mosaiced long-slit data were reduced in the way
described in Kniazev et al. (2008). The resulting rich
emission-line spectra of WS1 and WS2 are presented and discussed
in Section\,\ref{sec:spec}.

Equivalent widths (EWs), FWHMs and radial heliocentric velocities
(RVs) of main emission lines in the spectra of WS1 and WS2
(measured applying the MIDAS programs; see Kniazev et al. 2004 for
details) are summarized in Table\,\ref{tab:ws1} and
Table\,\ref{tab:ws2}, respectively.

We note that SALT is a telescope with a variable pupil, so that
the illuminating beam changes continuously during the
observations. This makes the absolute flux/magnitude calibration
impossible, even using spectrophotometric standard stars or
photometric standards.

\subsection{Photometry}
\label{sec:phot}

WS2 is indicated in the International Variable Star Index (VSX) as
a suspected variable with maximum visual magnitude of 12.5 (Watson
2006). Although the source of this magnitude was not indicated, we
found that it comes from the paper by Sanduleak \& Stephenson
(1973), who estimated $V\simeq 12.5$ mag as the average limiting
magnitude for detection of a continuum in their low-dispersion
spectra obtained with an objective prism (see also
Section\,\ref{sec:dis}). Since Sanduleak \& Stephenson (1973) did
not detect the continuum in the spectrum of WS2, the actual visual
brightness of this star at the moment of their observations might
have been fainter than 12.5 mag. The variable status of WS2,
however, can be supported by the comparison of the 2MASS and the
DENIS $J$ and $K_{\rm s}$ magnitudes (see Table\,\ref{tab:phot}),
which shows that in 1998 the star brightened in the $K_{\rm
s}$-band by $\simeq 0.84\pm 0.19$ mag during $\simeq 2.5$ months.

\begin{table*}
  \caption{Archival and contemporary photometry of WS1 and WS2.}
  \label{tab:phot}
  \renewcommand{\footnoterule}{}
    \begin{tabular}{lccccccccc}
      \hline
      &  &  & WS1 & & & \\
      \hline
      Date & $B$ & $V$ & $R$ & $I$ & $J$ & $K_{\rm s}$ \\
      \hline
      1976 March 8$^{(1)}$ & 17.50$\pm$0.10 & -- & -- & -- & -- & -- \\
      1979 June 7$^{(1)}$ & -- & -- & -- & 12.60$\pm$0.10 & -- & -- \\
      1984 February 22$^{(1)}$ & -- & -- & 14.10$\pm$0.10 & -- & -- & -- \\
      1993 June 20$^{(1)}$ & -- & -- & 14.30$\pm$0.10 & -- & -- & -- \\
      1998 July 1$^{(2)}$ & -- & -- & -- & 12.79$\pm$0.03 & 10.56$\pm$0.10 & 8.90$\pm$0.11 \\
      2000 February 22$^{(3)}$ & -- & -- & -- & -- & 10.33$\pm$0.02 & 8.90$\pm$0.02 \\
      2011 June 12$^{(4)}$ & -- & 15.25$\pm$0.04 & -- & -- & -- & -- \\
      2011 September 23$^{(5)}$ & 17.36$\pm$0.03 & 15.31$\pm$0.03 & 13.62$\pm$0.03 & 12.18$\pm$0.03 & -- & -- \\
      \hline
      &  &  & WS2 & & & \\
      \hline
      Date & $B$ & $V$ & $R$ & $I$ & $J$ & $K_{\rm s}$ \\
      \hline
      1958 April 17$^{(1)}$ & 17.70$\pm$0.10 & -- & -- & -- & -- & -- \\
      1976 May 31$^{(1)}$ & 16.90$\pm$0.10 & -- & -- & -- & -- & -- \\
      1980 May 1$^{(1)}$ & -- & -- & -- & 10.20$\pm$0.10 & -- & -- \\
      1984 April 9$^{(1)}$ & -- & -- & 12.20$\pm$0.10 & -- & -- & -- \\
      1991 August 5$^{(1)}$ & -- & -- & 12.70$\pm$0.10 & -- & -- & -- \\
      1998 August 10$^{(3)}$ & -- & -- & -- & -- & 6.62$\pm$0.04 & 4.70$\pm$0.02 \\
      1998 October 29$^{(2)}$ & -- & -- & -- & 9.86$\pm$0.05 & 6.18$\pm$0.12 & 3.86$\pm$0.19 \\
      2011 June 13$^{(4)}$ & -- & -- & -- & 10.13$\pm$0.04 & -- & -- & -- \\
      2011 September 10$^{(4)}$ & -- & 14.10$\pm$0.04 & -- & -- & -- & -- \\
      2011 September 20$^{(5)}$ & 17.80$\pm$0.03 & 14.57$\pm$0.03 & 11.89$\pm$0.03 & 9.72$\pm$0.03 & -- & -- \\
      \hline
      \MC{6}{l}{(1) USNO B-1 (Monet et al. 2003); (2) DENIS; (3) 2MASS; (4) SALT; (5) PROMPT.}
           \end{tabular}
     \end{table*}

To determine the contemporary photometry of WS1 and WS2, we
observed these stars with the PROMPT network, which consists of
six 0.4-m robotic telescopes operating in Chile by the University
of North Carolina (Reichart et al. 2005). Observations were
carried out on 2011 September 20 and 23 with the PROMPT No.\,4
telescope, which is equipped with $BVRI$ filters of the
Johnson-Cousins photometric system. Transformations from the
instrumental to the standard photometric system were obtained
during several nights in 2010 using Landolt (1992) standard stars
in the regions of RU 149, RU 152, and SA 98. The photometric
accuracy is $\simeq 0.03$ mag in all the filters. The results are
presented in Table\,\ref{tab:phot}, where we also give
photographic magnitudes obtained from different sources and
calibrated using the secondary photometric standards established
from the PROMPT data, $J$ and $K_{\rm s}$ magnitudes from 2MASS,
and $I,J$ and $K_{\rm s}$ magnitudes from the DENIS database (The
DENIS Consortium, 2005).

\begin{table}
  \caption{Details of central stars associated with two circular shells discovered with WISE.}
  \label{tab:det}
  \renewcommand{\footnoterule}{}
  \begin{minipage}{\textwidth}
    \begin{tabular}{cccccc}
      \hline
       & WS1 & WS2 (= Hen\,3-1383) \\
      \hline
      RA(J2000) &  $13^{\rm h} 36^{\rm m} 28\fs63$ & $17^{\rm h} 20^{\rm m} 31\fs76$ \\
      Dec.(J2000) &  $-63\degr 45\arcmin  38\farcs7$ & $-33\degr 09\arcmin 49\farcs0$ \\
      $l$ & 307.8856 & 353.5275 \\
      $b$ & -1.3230 & 2.2005 \\
      $B$ (mag) & 17.36$\pm$0.03 & 17.80$\pm$0.03 \\
      $V$ (mag) & 15.31$\pm$0.03 & 14.57$\pm$0.03 \\
      $[3.4]$ (mag) & 8.03$\pm$0.03 & 4.18$\pm$0.05 \\
      $[4.6]$ (mag) & 7.59$\pm$0.02 & 2.79$\pm$0.03 \\
      $[12]$ (mag) & 6.76$\pm$0.03 & 2.58$\pm$0.02 \\
      $[22]$ (mag) & 4.62$\pm$0.04 & 1.65$\pm$0.03 \\
      \hline
    \end{tabular}
    \end{minipage}
   $^{a}$PROMPT; $^{b}$2MASS; $^{c}$WISE.
    \end{table}

Table\,\ref{tab:phot} shows that WS1 brightened in the $I$-band by
$0.61\pm 0.04$ mag during the last 13 years. One can also see that
this brightening was accompanied by the brightness increase in the
$R$-band by $0.68\pm0.10$ mag. Thus, the star apparently became
redder during the last $13-18$ years. Similarly,
Table\,\ref{tab:phot} shows that WS2 became fainter in the
$V$-band by $0.47\pm0.05$ mag during 10 days of 2011 September,
and brightened in the $I$-band by $0.41\pm0.05$ mag during three
months since the first spectrum was taken, i.e. the star appears
to have significantly reddened on a time-scale of several months.
Table\,\ref{tab:phot} also shows that WS2 experienced significant
($0.5-0.9$ mag) variability in the $B$- and $R$-bands on
time-scales of decades, although the poor time cadence leaves the
possibility that the brightness changes in these bands occurred on
much shorter time-scales.

The details of WS1 and WS2 are summarized in Table\,\ref{tab:det},
where we give their 2MASS coordinates, the contemporary $B$ and
$V$ magnitudes, and, for the sake of completeness, the WISE
magnitudes from the WISE Preliminary Release Source
Catalog\footnote{http://irsa.ipac.caltech.edu/cgi-bin/Gator/nph-dd}.

\begin{figure*}
\begin{center}
\includegraphics[width=12cm,angle=270,clip=]{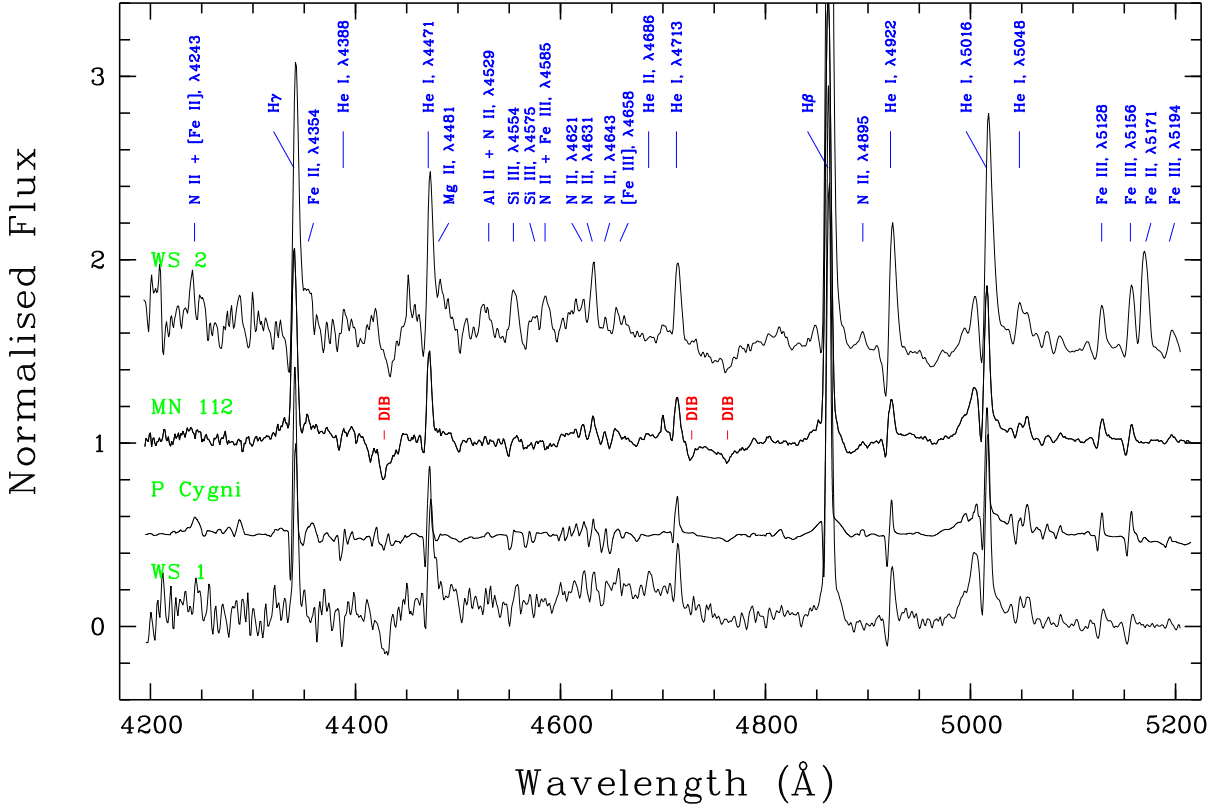}
\end{center}
\caption{Comparison of normalised blue spectra of WS1 and WS2 with
those of the prototype LBV P\,Cygni and cLBV MN112. The principal
lines and most prominent DIBs are identified.} \label{fig:blue}
\end{figure*}

\section{WS1 and WS2 -- new Galactic (candidate) LBV stars}
\label{sec:spec}

\begin{figure*}
\begin{center}
\includegraphics[width=12cm,angle=270,clip=]{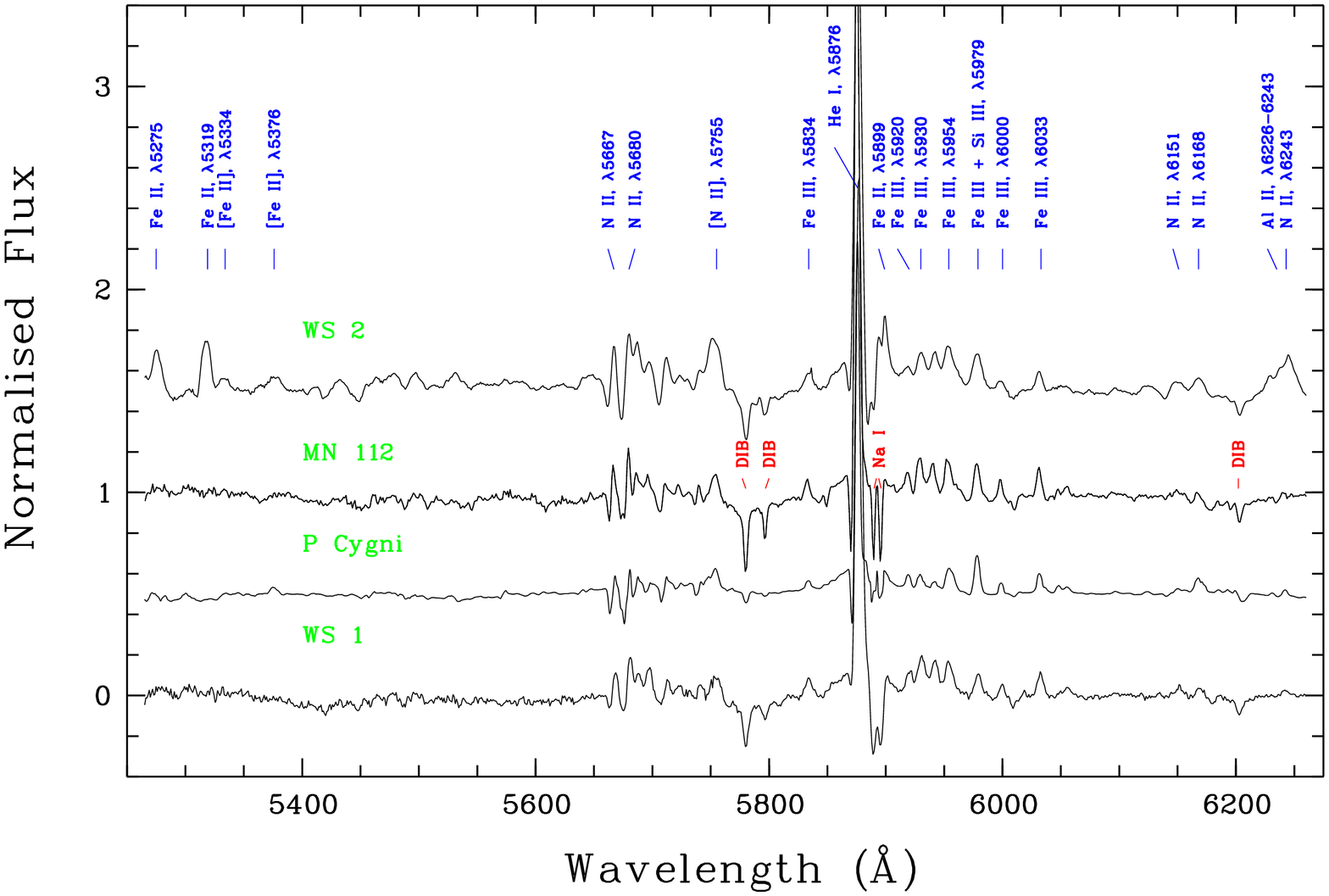}
\end{center}
\caption{Comparison of normalised green spectra of WS1 and WS2
along with spectra of the prototype LBV P\,Cygni and cLBV MN112.
The principal lines and most prominent DIBs are identified.}
\label{fig:green}
\end{figure*}
\begin{figure*}
\begin{center}
\includegraphics[width=12cm,angle=270,clip=]{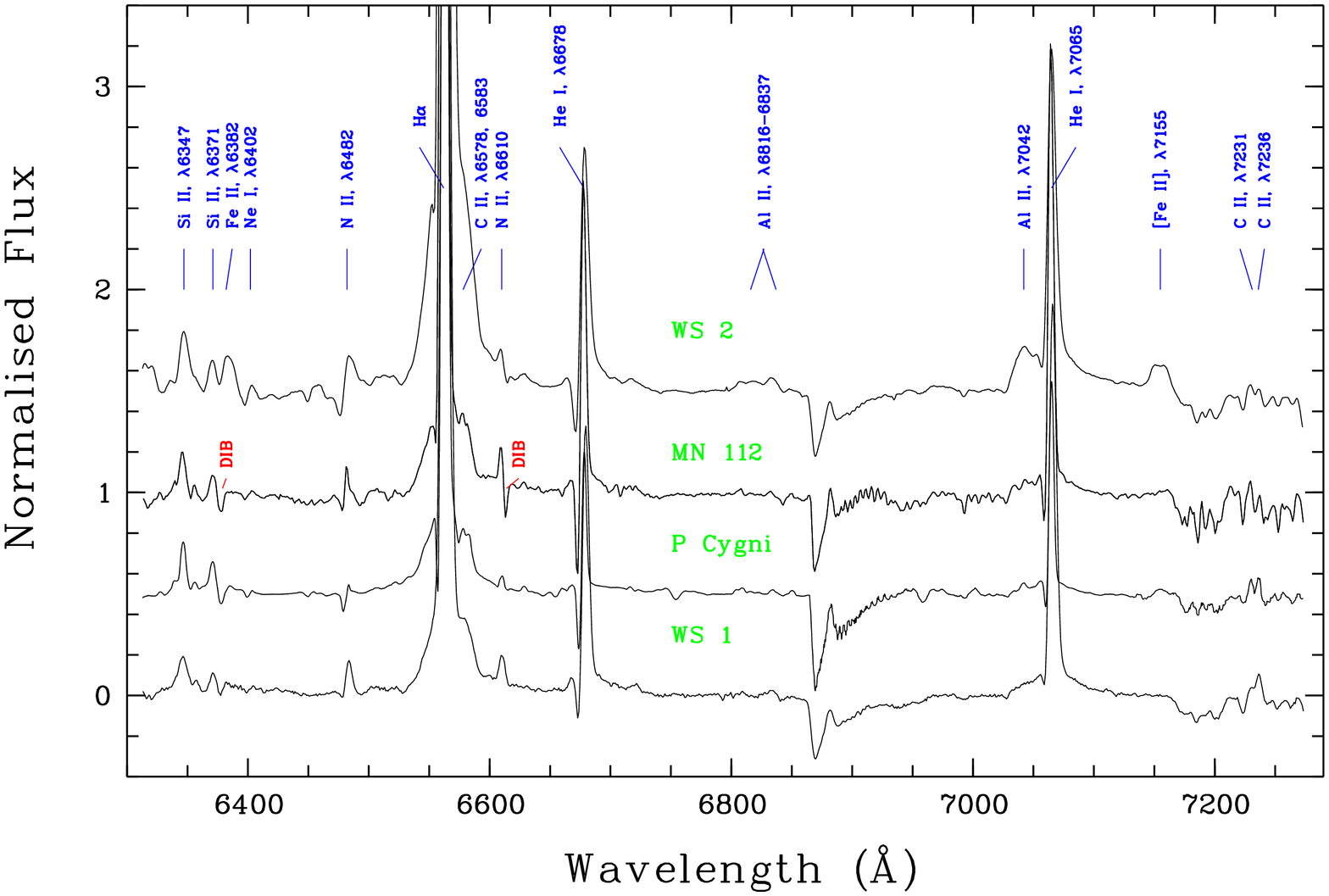}
\end{center}
\caption{Comparison of normalised red spectra of WS1 and WS2 along
with spectra of the prototype LBV P\,Cygni and cLBV MN112. The
principal lines and most prominent DIBs are identified.}
\label{fig:red}
\end{figure*}

Figures\,\ref{fig:blue}-\ref{fig:red} present the normalised
spectra of WS1 and WS2 (taken on 2011 June 12 and 13,
respectively) in the blue, green and red regions, where most of
the identified lines are marked. For the sake of comparison, we
also show in these figures the corresponding spectra of the
prototype LBV P\,Cygni (taken from Stahl et al. 1993 and degraded
to the resolution of our
spectra)\footnote{http://www.lsw.uni-heidelberg.de/users/ostahl/pcyg/}
and the recently discovered Galactic cLBV MN112 (Gvaramadze et al.
2010b). The spectra are arranged in such a way that objects with
the supposedly cooler effective temperature (see below) are placed
closer to the top of the figures. One can see that all four
spectra are very similar, which suggests the LBV classification
for WS1 and WS2 as well. To identify the lines indicated in the
figures, we mostly used the spectral atlas of P\,Cygni by Stahl et
al. (1993).

All four spectra are dominated by strong emission lines of H and
He\,{\sc i}; some of them show P\,Cygni profiles. Both H$\alpha$
and H$\beta$ have prominent wings. Other emission lines present in
all four spectra are numerous metal lines of N\,{\sc ii}, Fe\,{\sc
iii} and Si\,{\sc ii}, and the carbon doublets at $\lambda\lambda$
6578, 6583 and $\lambda\lambda$ 7231, 7236. The dichotomy of
Fe\,{\sc iii} lines observed in the spectra of P\,Cygni (Stahl et
al. 1993) and MN112 (Gvaramadze et al. 2010b) is present in the
spectra of WS1 and WS2 as well: lines with low multiplet numbers
(e.g. Fe\,{\sc iii} $\lambda\lambda 5127, 5156$) have distinct
P\,Cygni profiles, while those with higher multiplet numbers (e.g.
Fe\,{\sc iii} $\lambda\lambda 5920-6033$) are purely in emission.
Some of the N\,{\sc ii} lines (e.g. $\lambda\lambda 6151, 6168$)
are also purely in emission. The only forbidden line common for
all four spectra is that of [N\,{\sc ii}] $\lambda 5755$, although
the presence of weak [N\,{\sc ii}] $\lambda\lambda 6548, 6583$
lines in the H$\alpha$ emission wings cannot be excluded.

Numerous diffuse interstellar bands (DIBs) are present in the
spectra. In the blue region they are $\lambda\lambda 4428, 4726$
and 4762, in the green and red the strongest DIBs are at 5780,
5797, 6379 and 6613 \AA. Strong absorptions visible at $\lambda >
6800$ \AA \, are telluric. The sodium doublet in all four spectra
is of interstellar origin. In addition to the interstellar lines,
the spectrum of WS2 (as well as that of P\,Cygni; Stahl et al.
1993) shows blueshifted ($\Delta v \simeq -220 \, \kms$)
components of circumstellar origin.

There are, however, several important distinctions in the spectra.
The blue spectrum of WS1 shows the presence of a weak emission of
He\,{\sc ii} $\lambda 4686$ and the high-ionization forbidden line
[Fe\,{\sc iii}] $\lambda 4658$, which indicates that this star is
somewhat hotter than the other three stars. The presence of these
lines makes the spectrum of WS1 almost identical to that of the
bona fide LBV AG\,Car during the epoch of a minimum in 1985-1990,
when it was a WN11 star (Stahl et al. 2001) with an effective
temperature of $\simeq 22000-23000$ K (Groh et al. 2009a). Thus,
WS1 can be classified as a WN11 star. The WN11 classification of
WS1 also follows from the position of this star on the diagrams
showing a comparison of EWs of He\,{\sc i} $\lambda 5876$ versus
He\,{\sc ii} $\lambda 4686$ lines and He\,{\sc ii} $\lambda 4686$
EW versus FWHM for late WN-type (WNL) stars in the Milky Way and
the LMC (see Fig.\,1 of Crowther \& Smith 1997). With
EW(5876)=16.5 \AA, EW(4686)=0.4 \AA, and FWHM(4686)=10.8 \AA \,
(see Table\,\ref{tab:ws1}), WS1 lies in the region occupied by
WN11 stars. We conclude, therefore, that WS1 is a cLBV near a
brightness minimum.

The spectrum of WS2 shows several permitted and forbidden emission
lines of singly-ionized iron, which are absent or too weak to be
detected in the spectra of other three stars. Of these lines the
most prominent are Fe\,{\sc ii} $\lambda\lambda 5171, 5275, 5319,
5899, 6382$ and [Fe\,{\sc ii}] $\lambda\lambda 5334, 5376, 7155$.
Some of the Fe\,{\sc ii} lines ($\lambda\lambda 5892-5897$) are
blended with the blueshifted component of the Na\,{\sc I} D1 line,
while some others (e.g. $\lambda\lambda 4922, 5016$) are blended
with the He\,{\sc i} lines. The latter results in an anomalously
high intensity ratio of He\,{\sc i} $\lambda 4922$ to He\,{\sc i}
$\lambda 4471$ compared to the other three stars (cf. Hillier et
al. 1998). Several emissions of Al\,{\sc ii} are also detected in
the spectrum, of which the most prominent are at $\lambda\lambda
6226-6243$ and $\lambda 7042$. Overall, the spectrum of WS2 is
very similar to those of AG\,Car in 2003 January (see Fig.\,3 in
Groh et al. 2009a) and the cLBV HD\,316285 (Hillier et al. 1998).
Since the singly-ionized iron lines appeared in the spectrum of
AG\,Car two years after a visual minimum in 2001 (when the
effective temperature of AG\,Car decreased from $\simeq 17000$ to
14000 K; Groh et al. 2009a), one can conclude that WS2 is also
near the brightness minimum and currently is either at the onset
of the rise to the maximum or on the way to the minimum.

Figures\,\ref{fig:blue}-\ref{fig:red} show that the emission lines
of WS2 are much broader than those of WS1 (cf. also
Tables\,\,\ref{tab:ws1} and \ref{tab:ws2}) and the two other
stars, which might be caused by a higher wind velocity,
$v_\infty$, of WS2. Indeed, this inference is supported by
measurements of the FWHM of the [N\,{\sc ii}] $\lambda 5755$ line,
which can be used as a measure of $v_\infty$ (e.g. Crowther,
Hillier \& Smith 1995). After correction for instrumental width
($\simeq 4.74 \pm 0.05$ \AA), we found $v_\infty \simeq 640\pm 20
\, \kms$ for WS2 and $\simeq 350\pm 40 \, \kms$ for WS1. A similar
value of $v_\infty$ for WS2 also follows from the width of the
flat-topped lines [Fe\,{\sc ii}] $\lambda\lambda$ 5334, 5376, and
7155: the mean value of $v_\infty$ derived from these three lines
is $\simeq 690 \pm 15 \, \kms$. For comparison, the wind
velocities of P\,Cygni and MN112 are $\simeq 200 \, \kms$ (e.g.
Barlow et al. 1994; Najarro, Hillier \& Stahl 1997) and $\simeq
400 \, \kms$ (Gvaramadze et al. 2010b), respectively. Although the
high wind velocity inferred for WS2 is more typical of WN8-9h
stars (Hamann, Gr\"{a}fener \& Liermann 2006), we note that
similar velocities were also measured for the Homunculus nebula of
$\eta$\,Car (e.g. Allen \& Hillier 1993; Meaburn, Walsh \&
Wolstencroft 1993) and for the LBV-like component of the WR binary
HD\,5980 in the Small Magellanic Cloud during its slow wind
eruptive phase in 1994, when this component was transformed from a
WR star in a LBV (Koenigsberger et al. 1998). This transformation
was accompanied by drastic changes of the wind velocity, which
reduced from $\sim 3000$ to $500 \, \kms$ and after the eruption
increased to the initial value. Thus, it is possible that WS2 has
either experienced a giant eruption in the past, so that the
inferred high velocity corresponds to the polar outflow \`{a} la
the Homunculus nebula, or that it is caught shortly before or soon
after an LBV-like eruption, similar to that occurred in HD\,5980
in 1994. In the latter case, one can expect that WS2 will show
major changes in its spectrum in the near future.

\begin{figure}
\begin{center}
\includegraphics[width=6cm,angle=270,clip=]{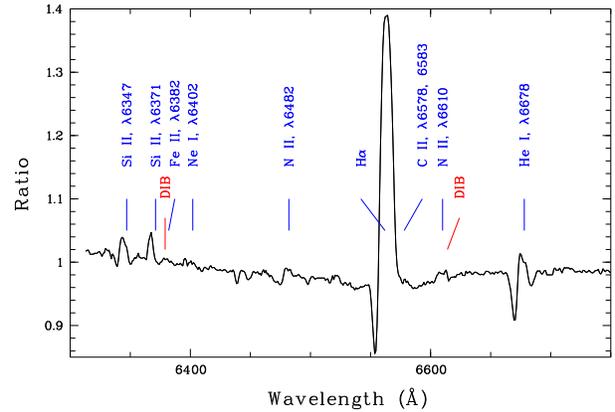}
\end{center}
\caption{The ratio of the normalised (red) spectra of WS2 obtained
within $\simeq 3$ months from each other. The positions of
principal lines and most prominent DIBs are indicated.}
\label{fig:comp-red}
\end{figure}

To search for possible spectral variability of WS2, we obtained a
second spectrum of this star $\simeq 3$ months after the first
observation. Table\,\ref{tab:ws2} shows that although most lines
remain invariable within their margins of error, some of them have
changed their characteristics significantly. The most noticeable
changes occurred in the EWs of the He\,{\sc i} $\lambda\lambda$
5016, 5876 lines (which became larger by $\simeq 30-40$ per cent)
and in the EW and the FWHM of the H$\alpha$ line (which decreased
and increased by $\simeq 20$ per cent, respectively). To
illustrate these changes graphically, we divided the first
normalised spectrum of WS2 by the second one.
Figs.\,\ref{fig:comp-green} and \ref{fig:comp-red} show the result
for two spectral ranges, $5700-6000$ \AA \, and $6300-6750$ \AA,
respectively. Changes in EWs, RVs and the line profiles manifest
themselves in the origin of artefacts above and below continuum.
Some of these artefacts show P\,Cygni-like profiles (e.g. at the
positions of the H$\alpha$ and He\,{\sc i} $\lambda 6678$ lines),
while others (e.g. He\,{\sc i} $\lambda 5876, 6678$) have an
additional inverse P\,Cygni-like profile.

Table\,\ref{tab:ws2} also shows that the FWHMs of the [N\,{\sc
ii}] $\lambda 5755$ and the flat-toped [Fe\,{\sc ii}] lines are
systematically reduced, which indicates the reduction of the wind
velocity as well. Using the FWHMs of these lines, we derived the
mean wind velocity of $\simeq 540\pm 20 \, \kms$, which implies
that $v_\infty$ decreased by $\simeq 100\pm 20 \, \kms$ during the
last three months (recall that this velocity decrease is
accompanied by the significant reddening of the star).

To conclude, the similarity between the spectra of WS1 and WS2 and
those of the well-known LBVs P\,Cygni and AG\,Car, along with the
significant short-term photometric variability and the presence of
the circumstellar shells strongly argue for the LBV classification
of both stars. For WS2 this classification is further supported by
the changes in the spectrum. We hope that spectroscopic and
photometric monitoring of WS1 and WS2 will allow us to confirm the
LBV status of these stars in the foreseeable future.

\section{Discussion}
\label{sec:dis}

\subsection{Classification of WS2}

\begin{figure}
\begin{center}
\includegraphics[width=6cm,angle=270,clip=]{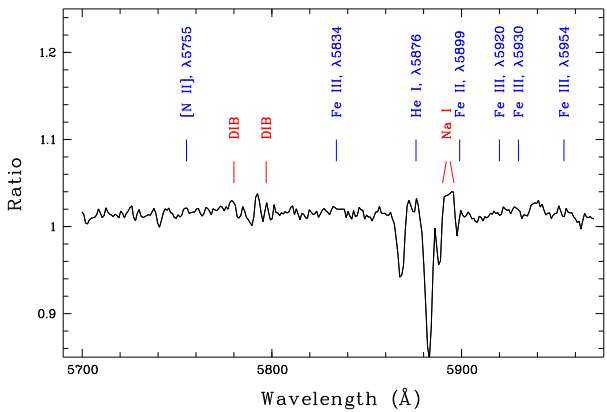}
\end{center}
\caption{The ratio of the normalised (green) spectra of WS2
obtained within $\simeq 3$ months from each other. The positions
of principal lines and most prominent DIBs are indicated.}
\label{fig:comp-green}
\end{figure}

We reported the detection of two circular shells with WISE and the
results of follow-up optical spectroscopic observations of their
central stars with SALT, which lead to the discovery of two new
Galactic cLBVs, WS1 and WS2. WS2 was known long ago as an
emission-line star Hen 3-1383 and was classified in the literature
as an M-type star, which is completely discrepant with our
classification of this star as a cLBV. Let us discuss the possible
origin of the discrepancy.

The existing spectral classifications of WS2 were based on the
observations carried out $\simeq 30-60$ years ago, so that one can
wonder whether WS2 has evolved blueward during the last several
decades? Although there are no known examples of such a radical
change in the spectral type of any know star, one cannot
completely exclude this possibility. For instance, in the recent
past WS2 might be a yellow hypergiant star, which, like the M33's
hypergiant star Var\,A (Humphreys, Jones \& Gehrz 1987),
experienced periodic changes of spectral type from pseudo M-type
to G-type during excursions on the cool side of the Yellow Void
(de Jager \& Nieuwenhuijzen 1997). These excursions were
accompanied by episodes of enhanced mass loss due to atmospheric
instability (de Jager 1998). Finally, several decades ago WS2 made
a last blueward swing and after crossing the Yellow Void found
themselves close to the S\,Doradus instability strip (Wolf 1989),
where we currently observe it as an LBV. WS2 therefore might be a
more evolved analog of the blueward evolving yellow hypergiant
IRC+10\,420 (Oudmaijer 1998), which most probably will turn into
an LBV in the close future (Humphreys, Davidson \& Smith 2002).
The characteristic timescale of formation and dispersal of
obscuring envelopes around yellow hypergiants ranges from years to
decades, which is comparable to the time period elapsed between
our and the previous spectroscopic observations of WS2.

The above considerations imply that WS2 has emerged as an LBV only
a few decades ago. This implication, however, is inconsistent with
the presence of the extended (parsec-scale; see below)
circumstellar shell around this star because shells of this scale
are rather typical of LBV and WNL stars, but were never observed
around known yellow hypergiants (e.g. Humphreys et al. 1997; Jura
\& Werner 1999; Smith et al. 2001; Lagadec et al. 2011). Although
one cannot exclude the possibility that WS2 became an LBV star
relatively long ago and that in the recent past it experienced a
giant eruption leading to the formation of an extended
pseudo-photosphere, which pushed the star on the cool side of the
Yellow Void (cf. Smith et al. 2004) and which dispersed during the
last several decades, we believe that the origin of the
discrepancy between the former and the present spectral
classifications of WS2 is rather more prosaic. Let us discuss the
sources of the previous classifications.

WS2 was identified as an emission-line star by The (1961), who
found a strong H$\alpha$ emission in its spectrum (obtained in
1960). The (1961) classified the star as M3 and suggested that "it
is possibly a long-period variable star", although no information
on the basis of these findings was provided. The visual brightness
of WS2 was measured to be 14 mag.

Sanduleak \& Stephenson (1973) classified WS2 as a Mep? star on
the basis of detection of an extremely strong H$\alpha$ emission
(with no evidence of H$\beta$ and He\,{\sc ii} $\lambda$4686
emissions) and the suspected presence of TiO bands. WS2 does not
show evidence of a continuum on the survey plates (taken in
1967-1968), from which Sanduleak \& Stephenson (1973) concluded
that the star was fainter than 12.5 mag at that time. This
magnitude is quoted in several subsequent catalogues (e.g. in
Watson 2006, where the star is indicated as a possible symbiotic
variable of the Z Andromedae type).

The TiO bands (and the continuum as well) were not visible in the
spectrum taken in 1949-1951 by Henize (1976), who interpreted the
presence of the very strong H$\alpha$ emission as the indication
that WS2 might be a symbiotic star.

The spectral classifications of WS2 given in the all three
above-mentioned papers were based on objective-prism spectra. The
first (low-dispersion) slit spectrum of WS2 was obtained in 1977
by Allen (1978), who suspected the presence of weak TiO absorption
and tentatively classified the star as a peculiar M-type
emission-line star (i.e. Mep?), because the H\,{\sc i} and
He\,{\sc i} emissions in its spectrum "are very strong for an Me
star".

To summarize, the existing historical spectroscopic data show that
WS2 is an emission-line star with the very strong H$\alpha$
emission (detected in all spectra starting from 1949 till 1977)
and the H\,{\sc i} emission (detected in 1977 in the first slit
spectrum of the star). TiO bands were suspected in 1967-1968 and
1978, and were not visible in 1949-1951. Although with this
information we cannot exclude the possibility that WS2 was an
pseudo M-type emission-line star $\simeq 30-60$ years ago, it is
plausible that the previous spectral classifications are erroneous
because of the low quality of the spectroscopic material,
aggravated by high visual extinction ($A_V \sim 10$ mag; see
below) towards the star.

\subsection{Reddening towards WS1 and WS2}

We now turn to estimate the reddening towards WS1 and WS2. First,
we matched the dereddened spectral slopes of these stars with
those of stars of similar effective temperature. Using the Stellar
Spectral Flux Library by Pickles (1998) and assuming the effective
temperatures of WS1 and WS2 of 22000 K and 14000 K (see
Section\,\ref{sec:spec}), respectively, we found the colour excess
$E(B-V)=2.40$ mag for WS1 and $E(B-V)=3.37$ mag for WS2. These
estimates only slightly depend on the assumed effective
temperature. By varying the temperature in a quite wide range
($\simeq 12000-26000$ K for WS1 and $\simeq 11000-23000$ K for
WS2), we found $E(B-V)\simeq 2.37\pm 0.10$ mag for WS1 and $E(B-V)
\simeq 3.36\pm 0.10$ mag for WS2. We also compared the spectral
slopes of WS1 and WS2 with that of AG\,Car (using the spectra of
AG\,Car obtained by us on 2009 June 1 and 2011 May 19), and found
that they nicely match each other if one dereddens the spectra of
WS1 and WS2, respectively, by $1.53-1.66$ and $2.61-2.73$ mag in
$E(B-V)$. Adopting $E(B-V)=0.63$ mag for AG\,Car (Humphreys et al.
1989), we found $E(B-V)=2.16-2.29$ mag for WS1 and
$E(B-V)=3.24-3.36$ mag for WS2, which are fairly well agree with
the above estimates.

The reddening can also be estimated by using the correlation
between the intensity of the DIBs and $E(B-V)$ (see Herbig 1995
for a review). In Table\,\ref{tab:dibs} we provide estimates of
$E(B-V)$ for both stars based on EWs of DIBs at $\lambda 5780$ and
$\lambda 5797$, and the relationships given in Herbig (1993). We
did not find this approach useful because of the large spread in
the resulting estimates of $E(B-V)$ and their inconsistency with
those derived from the spectral slopes. The discrepancy could be
caused by foreground regions of enhanced number density of
carriers of the DIBs. Another possibility is that the
circumstellar material around WS1 and WS2 produces the DIB
carriers and therethrough contributes to the strength of the
DIBs\footnote{We measured the EWs of DIBs in both spectra of WS2
and found that they agree with each other within the margin of
errors (see also Fig.\,\ref{fig:comp-green}).} (cf. T\"{u}g \&
Schmidt-Kaler 1981; Le Bertre \& Lequeux 1993; Heydari-Malayeri et
al. 1993). In what follows we assume $E(B-V)=2.4$ mag for WS1 and
$E(B-V)=3.4$ mag for WS2, and use these figures to constrain the
distances to and the luminosities of these stars. Using the
standard ratio of total to selective extinction $R_V =3.1$, we
obtained $A_V \simeq 7.4$ mag for WS1 and $A_V \simeq 10.5$ mag
for WS2.

\begin{table}
\caption{Estimates of the colour excess, $E(B-V)$, towards WS1 and
WS2 based on the equivalent widths (EWs) of diffuse interstellar
bands (DIBs) at $\lambda 5780$ and $\lambda 5797$ in the spectra
of these stars.} \label{tab:dibs}
\begin{center}
\begin{minipage}{\textwidth}
\begin{tabular}{ccccccc}
\hline
&  WS1 &  & WS2 \\
\hline
DIB & EW($\lambda$) & $E(B-V)$ & EW($\lambda$) & $E(B-V)$ \\
(\AA) & (\AA) & (mag) & (\AA) & (mag) \\
\hline
5780 & 1.79$\pm$0.22 & 3.38$\pm$0.52 & 1.78$\pm$0.12 & 3.36$\pm$0.46 \\
5797 & 0.83$\pm$0.19 & 6.06$\pm$0.88 & 0.91$\pm$0.12 & 6.69$\pm$0.85 \\
\hline
\end{tabular}
\end{minipage}
\end{center}
\end{table}

\subsection{Location of WS1 and WS2}

To estimate the distance to WS1 we use the empirical fact that
LBVs in the hot state are located on the S\,Doradus instability
strip (Wolf 1989). Assuming that the current effective temperature
of WS1 is $\simeq 22000-23000$ K (see Section\,\ref{sec:spec}), we
found the minimum luminosity of this star of $\log (L/L_{\odot})
\simeq 5.8$. Then assuming that the bolometric correction of WS1
is equal to that of AG\,Car during the epoch of minimum in
1985-1990, $-2.4$ mag (Groh et al. 2009a), we found the absolute
visual magnitude $M_V \simeq -7.3$ mag and the distance of $\simeq
11$ kpc (which places WS1 on the far side of the
Carina-Sagittarius arm). At this distance the linear radius of the
shell is $\simeq 1.8$ pc (i.e. a figure typical of LBV shells;
e.g. Weis 2001), while the distance from the Galactic plane is
$\simeq 240$ pc, which implies that WS1 is a runaway star (e.g.
Blaauw 1993). The luminosity of WS1 would be higher if this star
left the S\,Doradus instability strip and now undergoes an
outburst or recovers after it. Correspondingly, the distance to
the star would be larger, while its runaway status would be even
more unavoidable.

\begin{figure}
\begin{center}
\includegraphics[width=8cm,angle=0,clip=]{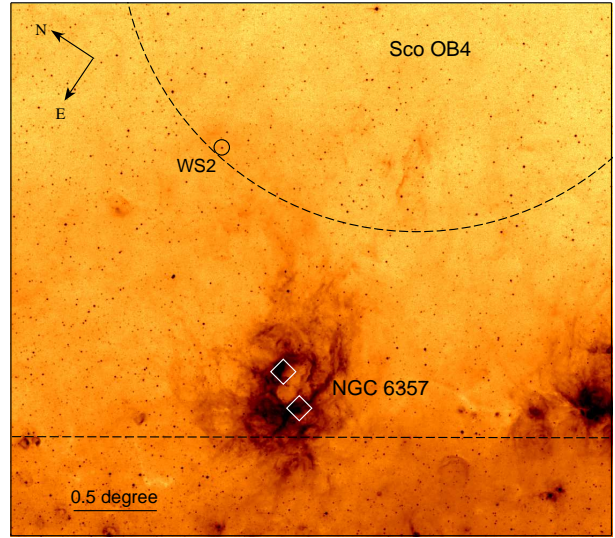}
\end{center}
\caption{{\it MSX} $8.3\,\mu$m image of the star-forming region
NGC\,6357 (containing the star clusters Pismis\,24 and
AH03\,J1725$-$34.4; indicated by diamonds) and its environments.
The position of WS2 is marked by a small circle. The approximate
boundary of the Sco\,OB4 association is shown by a large dashed
circle. The Galactic plane is shown by a dashed line.}
\label{fig:ScoOB4}
\end{figure}

In Section\,\ref{sec:shells}, we mentioned that WS2 was detected
not far ($\simeq 1.4$ degree to the northwest) from the
star-forming region NGC\,6357 and its associated massive star
clusters Pismis\,24 and AH03\,J1725$-$34.4, which are located on
the close side of the Carina-Sagittarius arm at a distance of
$\simeq 1.7$ kpc (Gvaramadze et al. 2011b). At the same time, WS2
is projected on the eastern periphery of the Sco\,OB4 association
(see Fig.\,\ref{fig:ScoOB4}), which is located at a distance of
$\simeq 1.1$ kpc (Kharchenko et al. 2005). Let us discuss whether
or not one of these stellar systems might be the birthplace of
WS2.

The possible association of the WC7 star WR\,93 with Pismis\,24
(Massey et al. 2001) implies that the age of this cluster is $\ga
2-2.5$ Myr, so that the cluster is old enough for some of its
(most massive) members to enter the LBV phase. On the other hand,
the $\simeq 40$ pc separation (in projection) of WS2 from
Pismis\,24 would imply that WS2 is a runaway star. Assuming that
WS2 was ejected from Pismis\,24 and that the ejection event
occurred $\sim 2$ Myr ago, one has the peculiar transverse
velocity of this star of $\simeq 20 \, \kms$. The peculiar
velocity of WS2, however, would be several times higher if the
star was ejected less than 2 Myr ago and if its space velocity has
a significant radial component. Such peculiar velocities can
easily be achieved in the course of three-body dynamical
encounters between very massive binary and single stars in the
dense cores of young star clusters (e.g. Gvaramadze \& Gualandris
2011). Detection of numerous very massive runaway stars around
very young ($\sim 1-2$ Myr) massive star clusters (Evans et al.
2010; Gvaramadze, Kroupa \& Pflamm-Altenburg 2010d; Bestenlehner
et al. 2011; Roman-Lopes, Barba \& Morrell 2011) supports the
theoretical expectation that the dynamical evolution of such
clusters is dominated by the most massive stars, so that the
ejected stars are also preferentially the most massive ones.
Unfortunately, the existing proper motion measurements for WS2 are
very unreliable, so that we did not attempt to calculate its
peculiar transverse velocity and thus to check its runaway status.

Assuming that the bolometric correction of WS2 is equal to that of
AG\,Car during the epoch of minimum in 2003, $-1.2$ mag (Groh et
al. 2009a), and adopting $V=14.1$ mag and the distance of 1.7 kpc,
we found $M_V \simeq -7.6$ mag and $\log (L/L_{\odot}) \simeq
5.4$\footnote{Note that these estimates should be considered
approximate because of the significant variability of WS2 (see
Table\,\ref{tab:phot}).}. These estimates imply that WS2 might
belong to a group of low-luminosity LBVs with $\log (L/L_{\odot})
\simeq 5.2-5.6$ (Humphreys \& Davidson 1994), which in turn would
imply that the progenitor of WS2 has evolved through a red
supergiant stage (Humphreys \& Davidson 1994; Smith et al. 2004).
The latter implication, however, means that the initial mass of
WS2 was $\la 25-30 \, \msun$, and that its age is $\ga 5$ Myr,
which is inconsistent with the young age of Pismis\,24, and argues
against the association between these two objects.

If WS2 is a member of the Sco\,OB4 association ($d=1.1$ kpc;
Kharchenko et al. 2005), then $M_V \simeq -6.6$ mag and $\log
(L/L_{\odot}) \simeq 5.1$. The resulting luminosity is therefore
well below the minimum luminosity found for any known (c)LBV and
would place WS2 on the hot side of the S\,Doradus instability
strip. Therefore, WS2 is most likely a background star projected
on Sco\,OB4 by chance.

It is also possible that WS2 is located in one of the next two
arms out, the Crux-Scutum Arm or the Norma Arm. In this case, the
distance to WS2 is either $\simeq 3$ or 5 kpc, which correspond to
$M_V \simeq -8.8$ mag and $\log(L/L_{\odot}) \simeq 5.9$ or $M_V
\simeq -9.9$ mag and $\log(L/L_{\odot}) \simeq 6.4$, respectively.
Note that for P\,Cygni and AG\,Car (during the minimum phase)
these parameters are equal, respectively, to $-8.0$ mag and 5.8
(Najarro et al. 1997) and $-8.1$ mag and 6.2 (Groh et al. 2009a).
At the distance of $3-5$ kpc the linear radius of the shell is
$\simeq 1.6-2.7$ pc, which is still within the range of radii
derived for LBV shells. Placing WS2 at a larger distance, say at 8
kpc, increases its $M_{\rm V}$ and $\log(L/L_{\odot}$) to $\simeq
-10.9$ mag and $\simeq 6.8$, respectively. This would make WS2 the
most luminous (c)LBV known to date, while its separation from the
Galactic plane and the radius of the shell would be $\simeq 300$
pc and $\simeq 4.4$ pc, respectively. Taken together, these
extreme figures make the large distances to WS2 unlikely. To
conclude, we deem that the distance to WS2 is either $\simeq 3$ or
5 kpc, although we cannot exclude the possibility that WS2 was
expelled from a star cluster (of appropriate age) hidden in the
molecular cloud associated with the star-forming region NGC\,6357
(cf. Gvaramadze et al. 2011b).

\subsection{WS1 and WS2 as runaway stars}

We now discuss the possibility that the LBV activity of WS1 and
WS2 and their possible runaway status might be related.

The fraction of runaway O~stars is found observationally to be
about 10-30 per cent (Gies 1987; Stone 1991; Zinnecker \& Yorke
2007). Gies (2007) summarizes the evidence for runaway O~stars to
often be rapid rotators. This idea is enhanced by the finding of
Walborn et al. (2010, 2011) that most of the spectroscopically
identified fast-rotating O~stars, the On~stars, are runaways or
are located in the field (and therefore most probably are runaways
as well; Gvaramadze et al. 2010d; Weidner et al. 2011).

Theoretically, a connection between the runaway status and fast
rotation of massive stars is expected. Mass transfer in a close
binary may spin up the mass receiver, which becomes a runaway star
upon the supernova explosion of the mass donor (Petrovic, Langer
\& van der Hucht 2005; Eldridge, Langer \& Tout 2011).
Alternatively, rapid rotators could also be produced in the course
of dynamical few-body encounters in young massive star clusters,
either because of merging of two (or several) stars or due to the
strong tidal interaction (Alexander \& Kumar 2001). In both cases,
the fast-rotating star could be ejected from the cluster (e.g.
Leonard 1995; Vanbeveren et al. 2009). Moreover, dynamical
encounters can also produce runaway binaries (e.g. Leonard \&
Duncan 1990; Kroupa 1998), which then can produce rapid rotators
by mass transfer or merging.

While the physical mechanism for LBV outbursts is still unclear,
there is consensus that the Eddington limit plays an important
role in these events (Lamers \& Fitzpatrick 1988; Smith \& Conti
2008) and that the critical rotational velocity of stars near
their Eddington limit becomes very small (Langer 1997, 1998). The
latter fact has two important consequences. First, it implies that
rotation must be important in LBV outbursts (Langer 1997; Langer
et al. 1999; Maeder \& Meynet 2000), which is supported by the
observation (e.g. Nota et al. 1995; Weis 2011) that the youngest
LBV nebulae have a bipolar shape (note that such nebulae may
become more spherical with time). Second, while fast rotation may
not be a sufficient condition to initiate the LBV phenomenon,
rapid rotators are more likely to become LBVs\footnote{Note that
the fast-rotating runaway stars might also be the progenitors of
long gamma-ray bursts (Hammer et al. 2006; Cantiello et al. 2007;
Eldridge, Langer \& Tout 2011; Le Floc'h et al. 2011).}. In stars
of relatively low initial mass, fast rotation may even trigger the
LBV activity, i.e. such stars would not become LBVs if they were
slow rotators.

In this connection, we note that tidal stripping of envelops of
evolved massive stars in close dynamical encounters might drive
stars towards the Eddington limit and thereby to trigger their LBV
activity. This would allow stars at the low-mass end of the range
of massive stars to manifest themselves as LBVs, even if they
cannot approach the Eddington limit in the course of an
unperturbed evolution (cf. Smith et al. 2011). Some of these stars
might also be ejected in the field.

Recently, Groh, Hillier \& Damineli (2006) and Groh et al. (2009b)
found that the bona fide LBVs AG\,Car and HR\,Car appear to rotate
close to their break-up velocities. Groh et al. (2009b) suggested
that the fast rotation is typical for all LBVs with strong
variability on short ($\sim$ decades) timescales, while the
dormant or ex-LBVs (e.g. P\,Cygni and HD\,168625) rotate more
slowly. These slowly-rotating (dormant/ex-) LBVs might have been
fast rotators in the past, which have lost a significant fraction
of their angular momentum due to one or several episodes of
violent mass ejection.

When rapid rotators are more likely to become LBVs, and rapid
rotators are often runaway stars, we conclude that LBVs may often
be runaway stars. This is supported by the observation that
AG\,Car and HR\,Car, like many other (c)LBVs, are located in the
field, i.e. far from known star clusters, and therefore are likely
runaways. We thus suggest that the LBV activity of WS1 and WS2
might be directly related to their runaway status.

\section{Summary and conclusion}
\label{sec:sum}

We have identified two new Galactic candidate LBVs via detection
of mid-infrared circular shells with WISE and follow-up
spectroscopy of their central stars with SALT. We have found that
the spectra of both stars are very similar to those of the
well-known LBVs P\,Cygni and AG\,Car (during its brightness minima
in 1985-1990 and 2003) and the recently discovered cLBV MN112. We
have also found that both stars, which we call WS1 and WS2, show
significant photometric variability. Namely, WS1 brightened in the
$R$- and $I$-bands by $0.68\pm0.10$ mag and $0.61\pm 0.04$ mag,
respectively, during the last 13-18 years, while WS2 (known as Hen
3-1383) varies its $B,V,R,I$ and $K_{\rm s}$ brightnesses by
$\simeq 0.5-0.9$ mag on time-scales from 10 days to decades. Taken
together, these findings strongly suggest that the detected
objects are LBVs. The LBV classification of WS2 is also supported
by the spectral variability of this star on a timescale of $\simeq
3$ months. To corroborate that WS1 and WS2 are bona fide LBVs, it
is necessary to detect the major changes in their brightness ($\ga
1-2$ mag) and spectra. Further spectroscopic and photometric
monitoring of these stars is therefore warranted. We have
discussed a connection between the location of massive stars in
the field and their fast rotation, and suggested that the LBV
activity of WS1 and WS2 might be directly related to their
possible runaway status.

To conclude, careful examination of the archival data of the WISE
survey and the {\it Spitzer} Legacy programs will undoubtedly lead
to new interesting discoveries, which hopefully will advance our
understanding the LBV phenomenon.

\section{Acknowledgements}

We are grateful to P. Morris (the referee) for his comments on the
manuscript. All SAAO and SALT co-authors acknowledge the support
from the National Research Foundation (NRF) of South Africa. This
work has made use of the NASA/IPAC Infrared Science Archive, which
is operated by the Jet Propulsion Laboratory, California Institute
of Technology, under contract with the National Aeronautics and
Space Administration, the SIMBAD database and the VizieR catalogue
access tool, both operated at CDS, Strasbourg, France.

\end{document}